\documentclass[prl, reprint]{revtex4-1}
\usepackage{amsmath,amsbsy, amssymb}
\usepackage[dvips]{graphicx}
\usepackage{xcolor}
\usepackage{units}
\DeclareGraphicsExtensions{.eps,.ps}

\usepackage{bm}

\def\Tr#1{{\textrm{ Tr}} \left( #1 \right)}
\def\Det#1{\textrm{Det}\left( #1 \right)}
\def\det#1{\textrm{Det}( #1 )} 

\def\ber{\begin{eqnarray}}
\def\eer{\end{eqnarray}}
\def\be{\begin{equation}}
\def\ee{\end{equation}}
\def\beno{\begin{equation*}}
\def\eeno{\end{equation*}}
\def\bea{\begin{eqnarray}}
\def\eea{\end{eqnarray}}

\begin{document}

\title{Peristaltic elastic instability in an inflated cylindrical channel}

\author{Nontawit Cheewaruangroj}
\affiliation{Cavendish Laboratory, University of Cambridge, 19 JJ Thomson Avenue, Cambridge CB3
0HE, United Kingdom}
\author{Karolis Leonavicius}
\affiliation{Department of Physiology Anatomy and Genetics, University of Oxford, Oxford OX1
3QX, UK}
\author{Shankar Srinivas}
\affiliation{Department of Physiology Anatomy and Genetics, University of Oxford, Oxford OX1
3QX, UK}
\author{John S. Biggins}
\affiliation{Cavendish Laboratory, University of Cambridge, 19 JJ Thomson Avenue, Cambridge CB3
0HE, United Kingdom}
\date{\today}
\begin{abstract}
A long cylindrical cavity through a soft solid forms a soft microfluidic channel, or models a vascular capillary. We observe experimentally that, when such a channel bears a pressurized fluid, it first dilates homogeneously, but then becomes unstable to a peristaltic elastic instability. We combine theory and numerics to fully characterize the instability in a channel through a bulk neo-Hookean solid, showing that instability occurs  supercritically with wavelength $2\pi/k=12.278....a$ when the pressure exceeds  $2.052....\mu$. In finite solids, the threshold pressure is reduced, and peristalsis is followed by a second instability which shears the peristaltic shape breaking axisymmetry. These instabilities shows that, counterintuitively, if a pipe runs through a bulk solid, the bulk solid can be destabilizing rather than stabilizing at high pressures. They also offers a route to fabricate periodically undulating channels, producing  waveguides with photonic/phononic stop bands.
\end{abstract}
 \maketitle

\begin{figure}[t]
       \includegraphics[width=0.95\columnwidth]{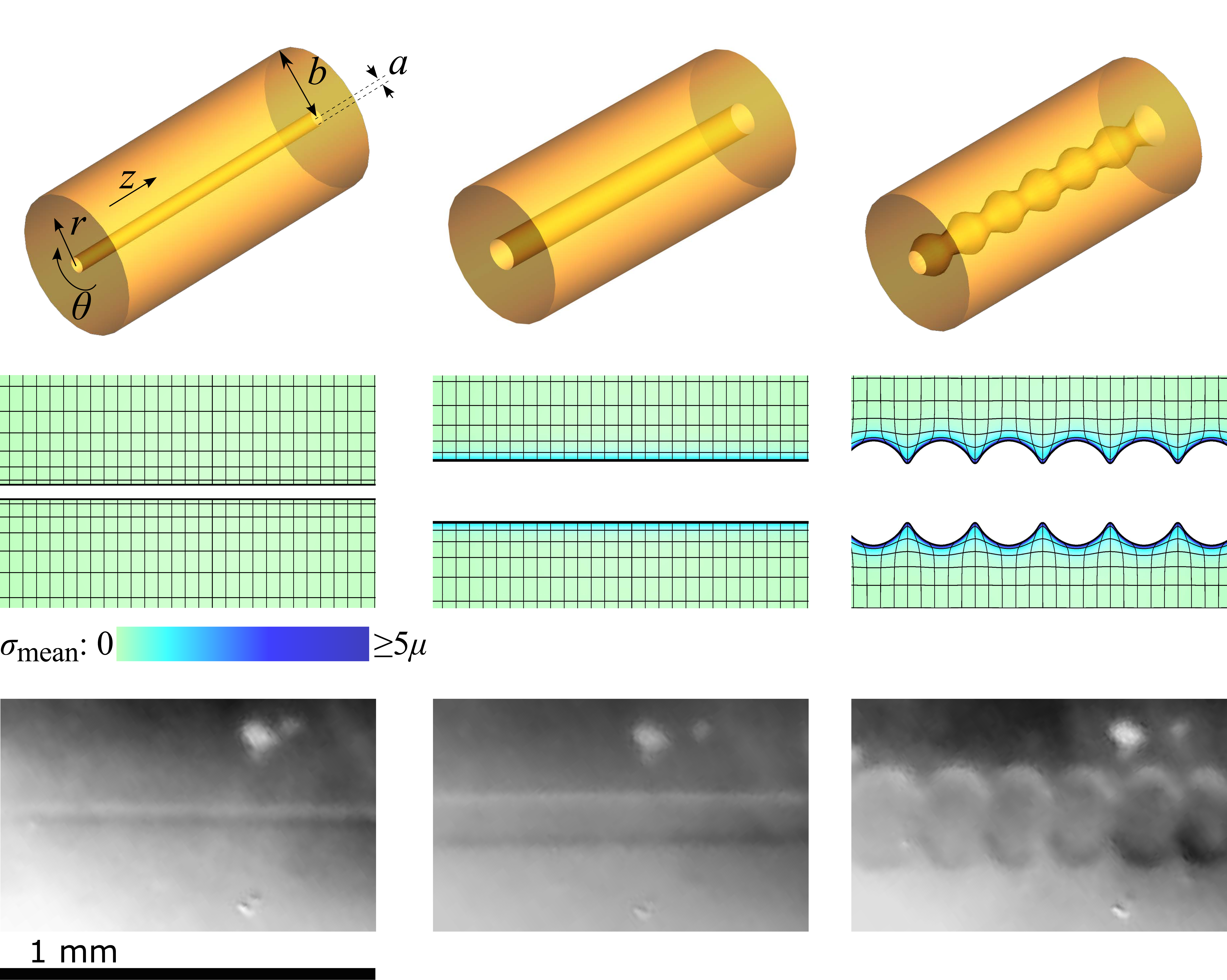}
\caption{Schematic (top), numerics (middle, neo-Hookean)  and experiment (bottom, polyacrymide) showing the shape evolution of a cylindrical channel through a soft solid under increasing internal pressure. At modest pressures the channel dilates simply, but at high pressures it  undergoes an elastic instability and adopts a peristaltic shape (right).} 
    \label{fig:peristaltic_pics}
\end{figure}

A fluid bearing channel through a soft solid is the prototypical element of biological plumbing \cite{pedley2000blood}, guiding fluid through the vascular, lymphatic, digestive, reproductive, renal and respiratory systems. Soft channels also underpin the blossoming field of soft microfluidics \cite{unger2000monolithic}, which exploits the convenience of soft lithography for rapid prototyping \cite{xia1998soft, mcdonalddt}, the deformability of soft channels to actuate valves and pumps \cite{unger2000monolithic, thorsen2002microfluidic, chou2001microfabricated}, and the mechanical compatibility between soft solids and soft tissues to build organs-on-chips and implantable clinical devices \cite{araci2014implantable, Koh366ra165}. Here we address a basic question about such channels: how do they change shape as their internal  pressure increases? As summarized in Fig.\ \ref{fig:peristaltic_pics}, we combine theory, numerics and experiment to show that, while  modest fluid pressures simply dilate such channels, when the pressure becomes comparable to the solid's shear modulus, the channel undergoes a reversible elastic instability and adopts a peristaltically undulating morphology.

The hallmark of soft solids, including elastomers, gels, and many biological tissues, is that they can sustain geometrically large elastic strains ($\gtrsim 100\%$) without failing, exposing them to the full complexity of large deformation geometry. This introduces a range of geometrically motivated elastic instabilities into their mechanics, including buckling \cite{euler1744additamentum}, creasing \cite{biotbook,trujillo2008creasing, cai2012creasing, hohlfeld2011unfolding, dervaux2012mechanical, tallinen2013surface} and wrinkling \cite{allen1969, cerda2003geometry} under compression; fingering \cite{shull2000fingering, saintyves2013bulk, biggins2013digital, mora2014gravity, biggins2015fluid}, fringing \cite{lin2016fringe, lin2017instabilities}  and beading \cite{mora2010capillarity, taffetani2015elastocapillarity, ciarletta2012peristaltic, xuan2017plateau} under tension; and ballooning \cite{mallock1891ii, alexander1971tensile, ogden1972large, gent1999elastic, meng2014phase}, aneurysm \cite{alexander1971tensile, gent2005elastic} and cavitation \cite{gent1959internal, ball1982discontinuous, cavitation} under inflation. These instabilities are important failure modes of soft/biological systems \cite{hirst1958dissecting}, but have also been exploited by evolution to sculpt developing brains \cite{brainpnas, brain}, guts \cite{loop, villi} and other organs \cite{dervaux2011shape, ciarletta2012growth, kucken2004model, diab2013ruga, razavi2016surface}, and by engineers to make shape-switching devices \cite{shim2012buckling, yang2015buckling, marthelot2017reversible, wang2018spatially}. The inflated channel instability we report here is a further example, with an analogous relationship to aneurysm as the sulcus instability has to buckling. 

We start by considering a cylindrical cavity, with initial radius $a$, running through soft elastic material initially occupying $a<r<b$, as shown in Fig.\ \ref{fig:peristaltic_pics}. If the channel is subject to an internal pressure $P_{in}$, it will dilate, deforming the solid and causing it to store elastic energy $E_{el}$. The observed dilation and deformation will be determined by the minimum of the effective energy
\begin{equation}
E_{tot} = E_{el}-P_{in} V,
\label{eq:etot}
\end{equation}
where $V$ is the dilated channel's volume. If the deformed solid has undergone a displacement field $\mathbf{u}$, its local shape change is described by the deformation gradient tensor $F=I+\nabla \mathbf{u}$, and its elastic energy density is of the form $W(F)$. We model the solid as an incompressible ($\mathrm{Det}\left({F}\right)=1$) neo-Hookean material \cite{Rivlin459} with shear modulus $\mu$, requiring  $W(F)=\frac{1}{2} \mu \Tr{F\cdot {F^T}}-P(\det{F}-1)$ where $P$ is a Lagrange multiplier pressure field enforcing incompressibility. This elastic model can be derived from the statistical mechanics of gaussian polymer networks \cite{wang1952statistical}, and therefore describes soft (i.e. lightly cross-linked) gels and elastomers well through to strains of several hundred percent \cite{biggins2013digital}. It is also the simplest elastic model capable of describing large deformations, and correspondingly offers the clearest exposition of geometrically motivated instabilities.

\begin{figure}[t]
       \includegraphics[width=0.9\columnwidth]{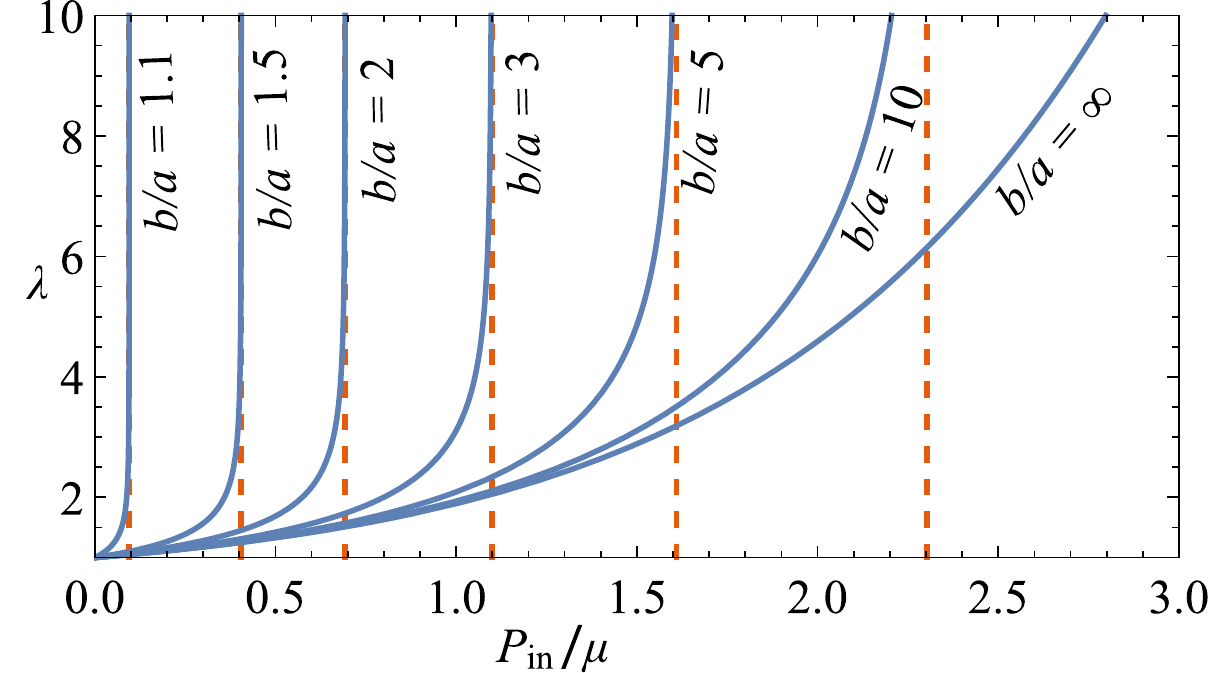}
\caption{Dilation factor, $\lambda$, of a cylindrical channel as a function of it's interior driving pressure $P_{in}$, for a range of aspect ratios $b/a$. The predicted dilation diverges at $P_{in}=\mu \log(b/a)$, a thick-walled balloon instability, but a channel through an bulk solid ($b\to\infty$) never diverges.}
    \label{fig:simple_dialation}
\end{figure}

Minimizing the total energy with respect to variations in $\mathbf{u}$ and $P$ gives the expected equations of mechanical equilibrium, and the constraint of incompressibility,
\begin{equation}
\nabla \cdot \sigma = 0, \mathrm{\ \ \ \ }\det{F} = 1,\label{eq:el2}
\end{equation}
where $\sigma=\frac{\partial W}{\partial F}= \left(\mu F - P F^{-T}\right),$ is the PK1 large deformation stress. These bulk equations are augmented by the natural inner and outer boundary conditions:
\begin{equation}
\left. (\sigma +P_{f}F^{-T})\cdot \hat{\mathbf{r}} \right|_{r= a,b} = 0,\label{bcs}
\end{equation}
where the boundary fluid pressure $P_f=P_{in},0$ at $r=a,b$ respectively. We first consider a simple dilation of the form $\mathbf{u}=u_0(r) \mathbf{\hat{r}}$, $P=\mu P_0(r)$. Incompressibility requires 
\begin{equation}
R\equiv r+u_0(r) =\sqrt{c^2+r^2},
\end{equation}
which generates a dilation of the the cavity radius by a factor of $\lambda=\sqrt{1+c^2/a^2}$. Mechanical equilibrium ($\nabla \cdot \sigma = 0$) then gives the form of the pressure field which, taking account of the stress free outer boundary, is
\begin{equation}
P_0=  \frac{1}{2} \left[\frac{r^2}{c^2+r^2}+\frac{b^2}{c^2+b^2}+\ln\left(\frac{b^2(c^2+r^2)}{r^2(c^2+b^2)} \right)\right]\notag.
\label{eq:q0}
\end{equation}
Finally, the inner boundary condition gives an implicit solution for the channel dilation, $\lambda$, which, introducing $g(x)\equiv (1/x)-\log(x)$, can be written
\begin{equation}
P_{in} = \frac{\mu}{2}\left[ g\left( 1+(a/b)^2(\lambda^2-1)\right)-g\left(\lambda^2\right) \right].\label{dial}
\end{equation}
This predicted dilation is plotted for a range of values of $a/b$ in Fig.\ \ref{fig:simple_dialation}. Dilation rises monotonically with $P_{in}$, and diverges at $P_{in}=\mu \log(b/a)$. If $b=a+t\approx a$, (a thin-walled pipe) this critical pressure reduces to $P_{in}\approx \mu t/a$, the signature of scaling of an aneurysm instability familiar to anyone who has inflated a party balloon. Conversely, in the $b\to\infty$ (bulk solid) limit eqn.\ (\ref{dial}) becomes 
\begin{equation}
P_{in} = \frac{\mu}{2}\left[ 1+\log(\lambda^2)-\lambda^{-2}\right],
\label{eq:cpcy} 
\end{equation}
showing that the channel dilates, but  only diverges at infinite pressure. In contrast,  the result for a spherical cavity, $P_{in} = \frac{\mu}{2}\left[ 5-4 \lambda^{-1}-\lambda^{-4} \right]$, diverges at $P_{in}=\frac{5}{2}\mu$, a celebrated result known as solid cavitation \cite{gent1959internal}.

We investigated this predicted stability of cylindrical cavities under inflation experimentally by cross-linking a 1mm thick rectangular slab of polyacrymide gel around a 30 $\mu$m wire (see experimental supplement for details). The wire was then removed (leaving behind a cylindrical channel), and the gel was equilibrated in a phosphate buffered saline solution (PBS). Finally, one end of the channel was plugged with a glass bead, and the channel was inflated by pumping additional PBS in through a glass capillary inserted at the other end. We used a computer controlled air pressure source to vary the pressure in the cavity, and monitored the evolving shape of the channel with a CMOS camera sensor fitted to a Leica stereomicroscope eyepiece. As seen in Fig.\ \ref{fig:peristaltic_pics} and supplementary video E1, we observed that the channel does indeed first simply dilate, in accordance with the above theory, but beyond a critical pressure it adopts a new peristaltically undulating morphology. The transition is  reversible, and occurs at quasi-static rates of inflation, suggesting a purely elastic mechanical instability.

   \begin{figure*}
\begin{center}
\includegraphics[width=180mm]{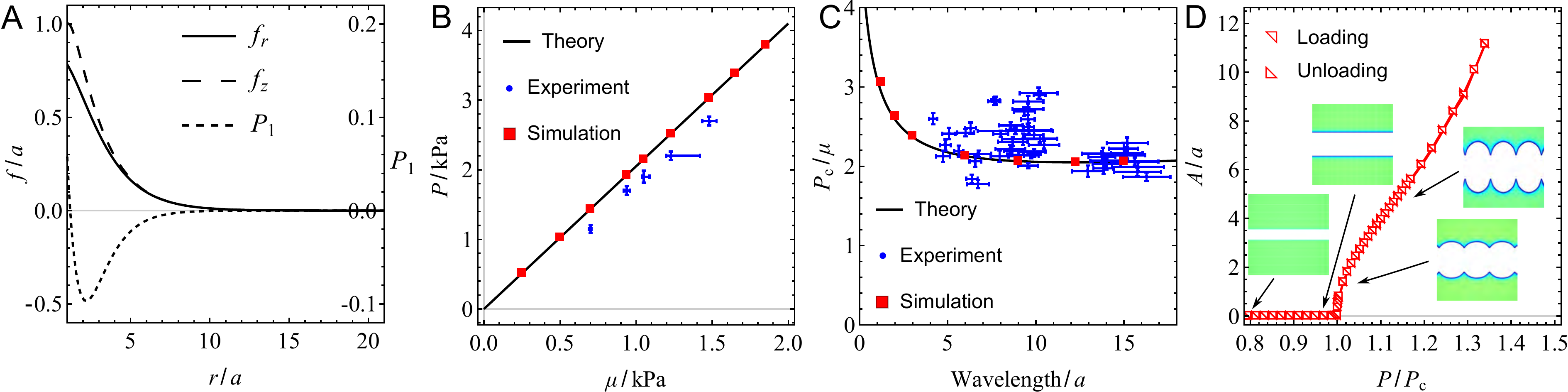}
\caption{Theoretical, numerical and experimental treatment of the peristaltic instability in a channel though a bulk solid. (a) Peristaltic fields for the first unstable mode from the stability analysis. (b) Critical pressure for instability as a function of shear modulus. (c) Theoretical/simulated critical pressure at each wavelength, compared with experimental wavelengths observed at and beyond threshold. (d) Simulated amplitude as a function of $P_{in}$, showing a supercritical transition without hysteresis.}\label{fig:peri_sol}
\end{center}
\end{figure*}

To understand this transition, we examine the stability of uniform dilation to small perturbations,  $\mathbf{u}=u_0\mathbf{\hat{r}}+\delta \mathbf{u}$, $P=\mu P_0+\delta P$, which, in turn, induce first order changes to the deformation gradient ($\delta F=\nabla \delta \mathbf{u}$) and the PK1 stress ($\delta \sigma=\mu \delta F-\delta P F_0^{-T}-\mu P_0 \delta(\Det{F}F^{-T})$). Expanding eqns.\ (\ref{eq:el2}-\ref{bcs}) to first order, we see the perturbation must satisfy $\nabla \cdot{\delta \sigma}=0$ (mechanical equilibrium), $\mathrm{Tr}(F_0^{-1}\cdot \delta F)=0$ (incompressibility) and $(\delta \sigma+P_{f} \delta(\Det{F}F^{-T}))\cdot \hat{\mathbf{r}}|_{r=a,b}=0$  (boundary conditions). Substituting a  peristaltic form 
\begin{align}
\delta\mathbf{u} &=   f_{r}(r) \cos(k z)\mathbf{\hat{r}}+   f_{z}(r) \sin(k z) \mathbf{\hat{z}} \nonumber \\ 
\delta P &=  \mu P_{1}(r) \cos(k z), \nonumber
\end{align}
into these equations (detailed algebra in the theoretical supplement) the condition of incompressibility reduces to 
\begin{equation}
R \left(R f_r'+k rf_z\right)+r f_r=0,\label{bulk_peturb}
\end{equation}
the equations of mechanical equilibrium become
\begin{align}
k f_r \left(r^2-R^2\right)^2+r R^3 \left(r f_z''+f_z'-k^2 r f_z+k rP_1\right)&=0\notag \\
r R^4 \left(r f_r''+f_r'-R P_1'\right)+r^2 f_r \left(r^2-R^2 \left(k^2 R^2+2\right)\right)& \\
+k Rf_z \left(r^2-R^2\right)^2&=0\notag
\end{align}
and the boundary conditions (at both $r=a,b$) become
\begin{align}
R f_z'-k r f_r&=0\notag \\
r R^2 f_r'-r^2 f_r-k r^2 R f_z-R^3 P_1&=0.\label{bc_perturb}\end{align}
%
%
Given $R\equiv r+u_0=\sqrt{a^2 \left(\lambda ^2-1\right)+r^2}$, equations (\ref{bulk_peturb}-\ref{bc_perturb}) form a fourth order generalized eigensystem for the critical degree of dilation, $\lambda$, required for instability. We solve the system numerically using matlab's bvp4c routine to find, for each $k$, the form of the perturbative fields and the threshold dialation. In Fig.\ \ref{fig:peri_sol} we summarize our results for a channel though a bulk ($b\to\infty$) solid. The first unstable mode occurs at $\lambda=4.824...$ (requiring $P_{in}=2.052...\mu$) and and with wavelength $2\pi/k=12.278...a$. We plot the form of this solution in Fig.\ \ref{fig:peri_sol}A, showing the fields take maximum values near the cavity, and decay  into the bulk over the length-scale $a$. Past threshold  a wide range of other wavelengths rapidly become unstable, as plotted in Fig.\ \ref{fig:peri_sol}C.

Across Fig.\ \ref{fig:peri_sol} we also compare these theoretical predictions with bespoke axisymetric finite element calculations (details in numerical supplement) and experiment. In particular, we show in Fig.\ \ref{fig:peri_sol}B that both numerical and experimental cavities indeed become unstable at $P_{in}=2.05...\mu$ over a range of gel moduli. This linearity in gel modulus is the hallmark of a purely elastic instability. In Fig.\ \ref{fig:peri_sol}C we show that finite element calculations with prescribed wavelengths indeed become unstable at the predicted pressure, and that, experimentally, a broad range of wavelengths are observed (a signature of the very flat theoretical wavelength-threshold curve) but that each wavelength is only seen around or above its predicted threshold pressure. Overall, the finite elements exactly reproduce the results of predicted thresholds and wavelengths, verifying our stability analysis, while the experimental data are in sufficient agreement to indicate our analysis has captured the essence of the observed instability. The experimental instability is often observed at pressures slightly below the theoretical value, which we attribute to the finite experimental slab thickness.

Stability analysis is limited to onset, but our finite element calculations can explore the peristaltic shape far beyond threshold. We conducted a numerical loading and unloading cycle in a bulk solid  (Fig.\ \ref{fig:peri_sol}D and supplementary video N1), which shows the amplitude growing and shrinking continuously above threshold without hysteresis; the instability is supercritical.

     \begin{figure*}
\begin{center}
\includegraphics[width=180mm]{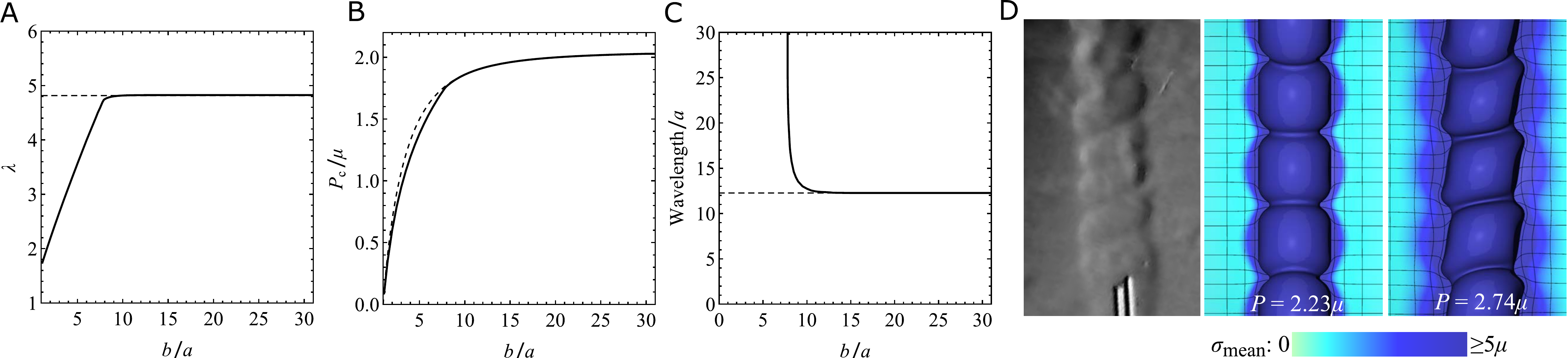}
\caption{Effect of finite outer radius on the peristaltic instability. (a-c) The critical dilation, pressure and wavelength of the first unstable mode. Dashed lines indicate the bulk solution, while solid lines indicate the full finite $b/a$ solution. (d) Experimental and numerical channels showing a second instability to a non axisymetric  sheared state.} 
\label{fig:finite_sol}
\end{center}
\end{figure*}

Finally, we analyze peristalsis at finite $b/a$. The stability equations (\ref{bulk_peturb}-\ref{bc_perturb}) only depend  on $P_{in}$ via the dilation it produces, and only depend on $b$ via the outer boundary condition. The instability thus has a universal form in all channels with even modestly large $b/a$, which all become unstable via the bulk solution, although the critical pressure is reduced in accordance with eqn.\ (\ref{dial}). In Fig.\ \ref{fig:finite_sol} we show the full form of the threshold dilation, pressure and wavelength as a function of $b/a$. As anticipated, these only deviate appreciably from the bulk form when $b/a\lesssim 10$, with wavelength growing and critical dilation falling in finite systems. We verify our stability analysis with finite-elements, which confirm (theory SI Fig.\ 1) our thresholds for many wavelengths and aspect ratios.


Experiments in thinner (0.5mm) slabs produced a peristaltic shape with a qualitatively new feature, an overall shearing that breaks axisymmetry, seen in Fig.\ \ref{fig:finite_sol}D. Inspired by this observation, we also conducted full 3-D finite element calculations for an idealized cylindrical channel, and observed such shearing to occur as a second instability (Fig.\ \ref{fig:finite_sol}D and videos N2 and N3) occurring at  $P_{in}\approx2.55...\mu$ for $b/a=10$. We reserve a full treatment of this instability for further work, but our initial results suggest the threshold diverges with $b$, explaining the lack of shearing in thick slabs. In thin-slab experiments shearing co-occurs with peristalsis because the rectilinear  slab breaks axisymmetry, encouraging the instability. 


In conclusion, we have shown that a cylindrical channel through a soft solid will spontaneously adopt a peristaltically undulating shape when pressurized to a multiple of the solid's shear modulus. The instability takes a simple form in channels through bulk solids, occurring at a critical pressure that is a simple multiple of the channel shear modulus, and with a wavelength which is a multiple of the cavity radius. These scalings are an inevitable consequence of the scale invariance of elasticity, which means channel radius is the only length-scale in the problem, and the elastic modulus is the only stress-scale. Bulk solid cavitation follows the same scale-free stress scaling, but peristalsis occurs at a lower pressure. 

\begin{figure}[b]
       \includegraphics[width=0.95\columnwidth]{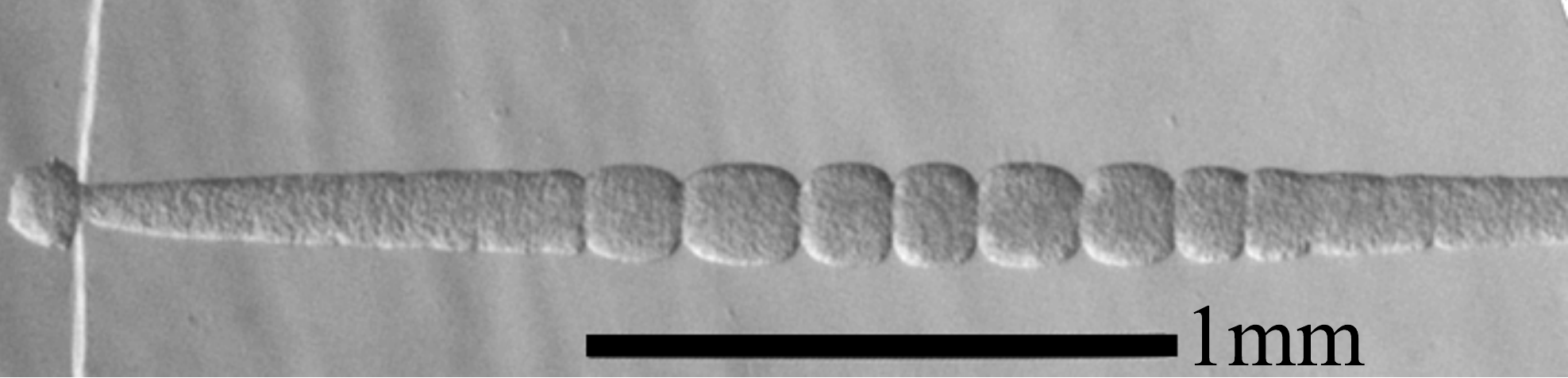}
\caption{Instability when a cell culture grows inside a channel.} 
    \label{fig:cells}
\end{figure}

The ultimate driver for peristalsis is that, for a given inflated channel volume, the peristaltic form requires less shape change in the surrounding solid and saves elastic energy. An instructive comparison is to the Plateau-Rayleigh instability \cite{rayleigh1878instability, mora2010capillarity, xuan2017plateau}, in which a cylinder subject to a surface energy transitions to a peristaltic shape to save surface area. Similarly here the peristalsis will reduce the surface area of the dilated channel \cite{xuan2016finite}, releasing stretch  energy. This geometric motivation suggests peristalsis will be universal in sufficiently deformable solids.  However, in less deformable elastic media, the large strain concentrations associated with peristalsis will precipitate fracture and failure. This suggests peristalsis places a fundamental limit on the pressure a channel can bear, just as Euler-buckling places a limit on the load of a column. Conversely, in sufficiently deformable solids, peristalsis offers a route to reversibly introduce periodicity into a channel. If implemented in a wave-guiding channel, this would allow a highly reflective photonic/phononic stop-band \cite{rayleigh1888xxvi, bertoldi2008mechanically} to be turned on and off.

More conventionally, the purpose of most biological/microfluidic channels is to guide flow. Fluid flows generate their own array of instabilities, and these are known to couple to to the elasticity of a deformable pipe. For example, the laminar/turbulent transition in a rigid pipe occurs at lower Reynolds numbers if the pipe is lined with a deformable solid \cite{lahav1973gel, krindel1979flow, ar2015stability}, and the lumen of such a ``gel-lined'' pipe responds with varicose and sinuous deformations  \cite{shankar2001asymptotic, shankar2009stability, ar2015stability}. However, the instability treated here is fundamentally different, narrowly because the driving dilation is excluded in treatments of gel lined pipes by their rigid outer wall \cite{ar2015stability}, and more broadly because peristalsis originates in solid rather than fluid mechanics. 

Many larger biological channels are better modelled as a thin walled stiff pipe running through a bulk soft solid. Under compression, thin-walled pipes buckle and collapse \cite{heil1997stokes, pedley2000blood}, as observed during the development of biological airways \cite{heil2008mechanics}. Under inflation, thin walled pipes fail via long-wavelength ballooning/aneurysm instabilities  \cite{mallock1891ii, alexander1971tensile, ogden1972large, gent1999elastic, meng2014phase, gent2005elastic}.  The addition of a surrounding bulk solid would naively be expected to suppress these instabilities, but here we show that surrounding soft tissue will become destabilizing  at sufficient inflation. An analogous situation arises when a stiff layer is adhered to a soft substrate and the entire ensemble is compressed. Compression results in surface wrinkling, which is essentially Euler-buckling in the stiff layer, constrained and suppressed by the soft substrate. However, if sufficient compression is applied, the substrate itself becomes unstable to the formation of cusped surface furrows known as sulci or creases:  the substrate becomes destabilizing rather than stabilizing. In this sense, peristalsis is a tensile analogue of the Biot instability, and the implied peristalsis-ballooning transition is likely to be a rich area for further study, just as the sulcus-wrinkle transition has been in compressive mechanics \cite{dervaux2011buckling, cao2012wrinkles, hutchinson2013role, budday2015period, wu2013swell, wang2014phase, guvendiren2010solvent, wang2015three, tallinen2015mechanics, auguste2017post}.

We first observed peristalsis in hydrogel channels containing growing mouse-embryo cultures (Fig.\ \ref{fig:cells}), where the dilation was driven by tissue growth over the course of weeks. Noting the biological ubiquity of the Biot instability, we speculate that the peristaltic instability will also manifest in biology, during both pathological and developmental processes. 


\begin{acknowledgments}
{\it Acknowledgements}. We thank T.\ Tallinen, on whose code our finite element calculations are based. N.C.\ thanks the Royal Thai Government Scholarship for funding.
\end{acknowledgments}
%

\end{document}